\documentclass[10pt]{iopart}
\usepackage{graphicx,iopams}
\begin{document}
\jl{12}
\submitted

\paper[Modeling of beam-customization devices for particle radiotherapy]{Computational modeling of beam-customization devices for heavy-charged-particle radiotherapy }
 
\author{Nobuyuki~Kanematsu$^{1,2}$, Shunsuke~Yonai$^{1}$, Azusa~Ishizaki$^{1,2}$ and Masami Torikoshi$^{1}$}
\address{$^1$\ Department of Accelerator and Medical Physics, Research Center for Charged Particle Therapy, National Institute of Radiological Sciences, 4-9-1 Anagawa, Inage-ku, Chiba 263-8555, Japan}
\address{$^2$\ Department of Quantum Science and Energy Engineering, Tohoku University, Aramaki-Aza-Aoba 01, Aoba-ku, Sendai 980-8579, Japan}

\eads{nkanemat@nirs.go.jp}

\begin{abstract}
A model for beam customization with collimators and a range-compensating filter based on the phase-space theory for beam transport is presented for dose distribution calculation in treatment planning of radiotherapy with protons and heavier ions.
Independent handling of pencil beams in conventional pencil-beam algorithms causes unphysical collimator-height dependence in the middle of large fields, which is resolved by the framework comprised of generation, transport, collimation, regeneration, range-compensation, and edge-sharpening processes with a matrix of pencil beams.
The model was verified to be consistent with measurement and analytic estimation at a submillimeter level in penumbra of individual collimators with a combinational-collimated carbon-ion beam. 
The model computation is fast, accurate, and readily applicable to pencil-beam algorithms in treatment planning with capability of combinational collimation to make best use of the beam-customization devices.
\end{abstract}


\section{Introduction}

In heavy-charged-particle radiotherapy with protons and heavier ions, conventional broad-beam systems deliver variety of volumetrically enlarged standard beams and an optimum one of them is chosen and customized to an individual treatment target (Kanematsu \etal 2007).
The beam customization is usually made by x-jaw, y-jaw, and multileaf collimators (XJC, YJC, and MLC) and custom-made accessories such as a patient collimator (PTC) and a range-compensating filter (RCF) with facility-specific variations.
For example, a PTC is always attached upstream of a RCF (Hong \etal 1996) or is optionally attached downstream of a RCF when the MLC field is not satisfactorily precise for the target (Kanai \etal 1999).
Despite inferiority in dose conformity, adaptiveness, and cost and labor with the accessories, the broad-beam delivery systems have clinical advantages in robustness against organ motion and in practicality of quality assurance over dynamic beam-scanning systems (Lomax \etal 2001 and J\"{a}kel \etal 2001), which will remain the same in the foreseeable future.

\begin{figure}
\begin{indented}
\item[] \includegraphics[width=10 cm]{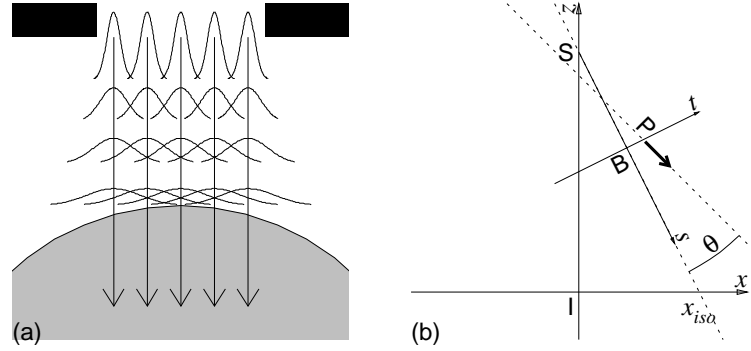}
\end{indented}
\caption{Side views of a beam field showing (a) the conventional PB model, where uniformly generated pencil beams (arrows) generated at a collimator (black) travel with growing spreads (curves) to enter a patient (gray), and (b) definitions of the broad beam's global coordinate $(z,x)$ and pencil beam's local coordinate $(s,t)$ systems, angle $\theta$, and position $x_\mathrm{iso}$, where S, I, B, and P stand for source, isocenter, pencil beam, and particle, respectively.}
\label{fig_1}
\end{figure}

In treatment planning of heavy-charged-particle radiotherapy, the pencil-beam (PB) algorithm has been commonly used for dose distribution calculations (Petti 1992, Hong \etal 1996, Deasy 1998, Kanematsu \etal 1998, Russel 2000, and Szymanowski 2001).
The PB algorithm handles a therapeutic broad beam as a set of independent pencil beams and superposes their doses to reproduce dose fluctuation from scatter in the presence of heterogeneity.
The pencil beams generated with certain angular spread at the collimator plane will grow spatially as they travel downstream side by side as shown in \fref{fig_1}(a).

Hong \etal (1996) formulated the PB total spread at a point of interest (POI) for a customized beam, which is represented using the standard gantry coordinate system (IEC 2002-3) as 
\begin{eqnarray}
\sigma_\mathrm{tot}^2 = \left( \frac{\sigma_\mathrm{src}} {z_\mathrm{col}-z_\mathrm{src}} \right)^2 \left( z-z_\mathrm{col}\right)^2 + \sigma_\mathrm{rcf}^2 + \sigma_\mathrm{pt}^2, \label{eq_1}
\end{eqnarray}
where the first term quadratically adds the spatial spread from the angular spread of the source size viewed from the collimator, $\sigma_\mathrm{src}/|z_\mathrm{col}-z_\mathrm{src}|$, in travel of the collimator--POI distance, $|z-z_\mathrm{col}|$, and the second and third terms add subsequent scatters from a RCF and a patient, respectively.
Their model is not exactly valid for a configuration with a PTC downstream of a RCF, where the constant spread $\sigma_\mathrm{rcf}$ from the RCF would cause an unphysical penumbra at the field edge even when the POI is in the proximity of the PTC.
Kanematsu \etal (1998) proposed an approximate model such that the scattering by the RCF is handled only in the angular spread for penumbra accuracy, ignoring the spatial spread in relatively short transport to the PTC, which may be formulated as
\begin{eqnarray}
\sigma_\mathrm{tot}^2 = \left[ \left( \frac{\sigma_\mathrm{src}} {z_\mathrm{col}-z_\mathrm{src}} \right)^2 + \theta_\mathrm{rcf}^2 \right] \left( z-z_\mathrm{col}\right)^2 + \sigma_\mathrm{pt}^2, \label{eq_2}
\end{eqnarray}
where $\theta_\mathrm{rcf}$ is the angular spread of the scatter from the RCF at height $z_\mathrm{rcf}$ and is related to the spread at the POI by $\sigma_\mathrm{rcf} = |z-z_\mathrm{rcf}|\, \theta_\mathrm{rcf}$ in \eref{eq_1}.

Those models were extensively examined against measurements in lateral penumbra of field edges (Hong \etal 1996, Kohno \etal 2004a, Akagi \etal 2006) though there has been little quantitative discussion on behaviors other than penumbra mainly due to dosimetric difficulties in the presence of heterogeneity (Kohno \etal 2004b).
In fact, both \eref{eq_1} and \eref{eq_2} should be inaccurate because the collimator height $z_\mathrm{col}$ would always affect the total spread $\sigma_\mathrm{tot}$ even in the middle of large fields, where the collimation should never be physically effective at all.
The inaccuracy is originated in the model where the pencil beams continue growing in the transport from the collimator regardless of its influence. 
Consequently, the PB size could be much larger than the heterogeneity, which would invalidate the beam model where all the involved particles are assumed to receive the same interactions.

In addition, those models can not handle combinational collimation with multiple collimators. 
In carbon-ion radiotherapy at National Institute of Radiological Sciences (Kanai \etal 1999), tumors longer than the available field length are treated with two unidirectional beams involving large patient movement between the deliveries, where the two fields are gently patched with the field edges by an upstream collimator for robustness against the setup errors while the other outline edges are sharply formed by the downstream collimator.
Besides, the upstream collimators should be optimized to minimize unwanted secondary radiations when the final collimator is not sufficiently thick. 
In the common practice, however, the combinational collimation is not fully utilized nor accurately handled in treatment planning.

In electron radiotherapy, the phase-space theory has been rigorously applied to the pencil-beam algorithm to deal with large scatter of electrons (Storchi and Huizenga 1985, Shiu and Hogstrom 1991, Boyd \etal 2001, and Chi \etal 2005).
Their method, the pencil-beam redefinition algorithm, effectively restricts the size of pencil beams and improves the accuracy against the heterogeneity.
The same idea should be applicable to heavy-charged-particle beams.

In this work, we develop an accurate computational model for customization of heavy charged beams based on the phase-space theory in analogy with the pencil-beam redefinition algorithm for electron beams, examine the model computation against measurement and analytic estimation with an example of a customized carbon-ion beam, and discuss the usability in practical treatment planning.

\section{Materials and methods}

\subsection{Physical models}\label{sec_physics}

In the PB algorithm, a subset of particles that occupy small area in the position--angle phase space are handled altogether as a Gaussian pencil beam (Kanematsu \etal 2006).
The development of the pencil beam is described by the Fermi-Eyges theory (Eyges 1948, Tomura \etal 1998, and Hollmark \etal 2004) with phase-space distribution of the involved $N$ particles,
\begin{eqnarray}
F(\theta, t) = \frac{N}{2\, \pi} \left(\overline{\theta^2}\, \overline{t^2} -\overline{\theta t}^2 \right)^{-\frac{1}{2}} \rme^{-\frac{\overline{t^2}\, \theta^2 -2\, \overline{\theta t}\, \theta\, t +\overline{\theta^2}\,t^2} {2\,\left(\overline{\theta^2}\,\overline{t^2} -\overline{\theta t}^2\right)}}, \label{eq_3}
\end{eqnarray}
where spatial variance $\overline{t^2}$, angular variance $\overline{\theta^2}$, and angular-spatial covariance $\overline{\theta t}$ are defined statistically with projected transverse position $t$ and angle $\theta$ as shown in \fref{fig_1}(b).
These phase-space parameters at the current position $s$ are related to the source size $\sigma_\mathrm{src}$ and the source distance $(s-s_\mathrm{src})$, or the focal-spot size and the focal distance in radiographic terminology, as
\begin{eqnarray}
\overline{\theta^2} = \frac{\sigma_\mathrm{src}^2}{(s-s_\mathrm{src})^2}, \qquad
\frac{\overline{t^2}}{\overline{\theta t}} = s-s_\mathrm{src}, \label{eq_4} \qquad
\frac{z-z_\mathrm{src}}{s-s_\mathrm{src}} = \vec{e_\mathrm{z}} \cdot \vec{e_\mathrm{s}},
\end{eqnarray}
where $s_\mathrm{src}$ is the $s$ coordinate of the source and $\vec{e_\mathrm{z}} \cdot \vec{e_\mathrm{s}} \approx -1$ is the scalar product of the basis vectors of the $z$ and $s$ axes that are nearly opposing in treatment systems.

In transport of a pencil beam, the angular variance propagates to the covariance and the spatial variance with increases
\begin{eqnarray}
\Delta \overline{\theta t} = \overline{\theta^2}\, \Delta s, \qquad
\Delta \overline{t^2} = \left(2\, \overline{\theta t} + \overline{\theta^2}\, \Delta s \right) \Delta s
\end{eqnarray}
in distance step $\Delta s$ while $\overline{\theta^2}$ stays constant in the transport.
Interactions with the air are ignored because of the small density ($\approx10^{-3}$ g/cm$^3$) while the effects are cared by experimental determination of the source size $\sigma_\mathrm{src}$ and the initial residual range $R_0$.

In matter of effective density $\rho$ (Kanematsu \etal 2003), the residual range $R$ and the angular variance $\overline{\theta^2}$ are modified in step $\Delta s$ as
\begin{eqnarray}
\Delta R = -\rho\, \Delta s, \qquad
\Delta \overline{\theta^2} = \overline{\mathcal{T}}\, \Delta s, \label{eq_6}
\end{eqnarray}
where $\overline{\mathcal{T}}$ is the mean of projected scattering power $\mathcal{T} = \rmd \overline{\theta^2}/\rmd s$ for multiple scattering, for the step. 
While the original Fermi-Eyges theory employed Rossi's formulation for multiple scattering (Eyges 1948), Highland (1975) introduced a logarithmic correction term for non-stochastic influence of single scattering, which was further generalized and extensively tested by Gottschalk \etal (1993). 

A differential form of the Highland-Gottschalk formula gives the scattering power for a particle with charge $Z\,e$, mass $A\,u$, and kinetic energy $E$ as
\begin{eqnarray}
\mathcal{T} = \left(1+\frac{1}{9}\, \log_{10} \ell \right)^2 \left(\frac{14.1\, \mathrm{MeV}}{pv}\, Z\right)^2\, \frac{1}{X_0}, \qquad
\ell = \int_{s_\mathrm{src}}^s \frac{\rmd s'}{X_0}, \label{eq_7}
\end{eqnarray}
where the momentum-velocity product $pv = E\,(E+2\, A\, u)/(E+A\, u)$ is quantified with the range--energy relationship $R = R(Z, A, E) \approx (A/Z^2)\,R(1,1,E/A)$ (ICRU 1993) and the normalized radiation length $\ell$ is the distance measured in the material-specific radiation length $X_0$ cumulatively along the beam down to the current point.

Use of geometric mean $\overline{pv} = \sqrt{p_0 v_0\, p_1 v_1}$ of the values before ($p_0 v_0$) and after ($p_1 v_1$) the step for the $pv$ in \eref{eq_7} leads to the mean scattering power $\overline{\mathcal{T}}$ for the step $\Delta s$ at a precision of 3\% or better for $\rho\, \Delta s < 0.7 R$ (Gottschalk \etal 1993).
Steps thicker than that should be recursively subdivided until the absolute error becomes small enough.
While the effective scattering point for the step is ideally $\Delta s/\sqrt{3}$ upstream of the step end (Yao \etal 2006), the concurrent energy loss moves the point slightly downstream, approximately at the center of the step (Gottschalk \etal 1993).

Loss of primary particles and yield of secondary particles in nuclear interactions are implicitly and approximately involved in the depth--dose curve being referenced in dose calculation.
This framework of the phase-space theory for charged-particle beams interacting with matter is also applicable, and in fact has been applied, to dose-distribution calculations in patient body with fine steps to deal with the heterogeneity (Kanematsu \etal 1998, 2006, and 2008 and Akagi \etal 2006).

\subsection{Beam development models}\label{sec_development}

\subsubsection{Generation }

A matrix of pencil beams is generated on a plane at height $z = z_0$ in the global coordinate system, where a broad beam is first effectively modified to form a lateral structure by either a collimator or a RCF.
The pencil beams are defined to have traveling distance $s = 0$ in the local coordinate systems shown in \fref{fig_1}(b).
They are placed at grid points $({x_0}_j,{y_0}_i)$ for row $i$ column $j$ with spacing $\delta_0$, which coincide with the field grids $({x_\mathrm{iso}}_j, {y_\mathrm{iso}}_i)$ with spacing $\delta_\mathrm{iso}$ on the isocenter plane in the beam's eye view, and are comprised of $N$ particles with residual range $R$ and slopes $a_\mathrm{x} = \rmd x/ \rmd z$ and $a_\mathrm{y} = \rmd y/ \rmd z$.
These parameters are handled as a function of height $z$ in transport with initial values
\begin{eqnarray}
\fl
\overline{x}(z_0)_{ij} = {x_0}_j, \quad
\overline{y}(z_0)_{ij} = {y_0}_i, \qquad
\overline{a_\mathrm{x}}(z_0)_{ij} = \frac{{x_0}_j}{z_0-z_\mathrm{src}}, \quad
\overline{a_\mathrm{y}}(z_0)_{ij} = \frac{{y_0}_i}{z_0-z_\mathrm{src}}, \nonumber\\*
\fl
R(z_0)_{ij} = R_0, \quad
\frac{N(z_0)_{ij}}{\delta_\mathrm{iso}^2} = \Phi_\mathrm{iso}\left({x_\mathrm{iso}}_j, {y_\mathrm{iso}}_i \right), \quad
\frac{\delta_0}{\delta_\mathrm{iso}} = \frac{{x_0}_j}{{x_\mathrm{iso}}_j} = \frac{{y_0}_i}{{y_\mathrm{iso}}_i} = \frac{z_0-z_\mathrm{src}}{z_\mathrm{iso}-z_\mathrm{src}},
\end{eqnarray}
where the relative number of particles $N$ reflects the original broad beam fluence expected on the isocenter plane, $\Phi_\mathrm{iso}$.
For the pencil beams generated with uniform distribution in square area $\delta_0^2$ and with the preserved focal distance, the phase-space parameters are initialized to
\begin{eqnarray}
\fl
\overline{\theta^2}(z_0)_{ij} = \frac{\sigma_\mathrm{src}^2}{(z_0-z_\mathrm{src})^2}, \qquad
\overline{\theta t}(z_0)_{ij} = \frac{\delta_0^2}{12\, |z_0-z_\mathrm{src}|}, \qquad
\overline{t^2}(z_0)_{ij} = \frac{\delta_0^2}{12},
\end{eqnarray}
where the source size $\sigma_\mathrm{src}$ reflects the effects of ignored scattering materials in the system as well as the initial beam emittance.
The beam field is conveniently represented with matrices of the PB parameters, $\bi{x}$, $\bi{y}$, $\bi{a_\mathrm{x}}$, $\bi{a_\mathrm{y}}$, $\bi{R}$, $\bi{N}$, $\overline{\btheta^2}$, $\overline{\btheta t}$, and $\overline{\bi{t}^2}$, that will develop in transport as a function of height $z$.

\subsubsection{Transport }\label{sec_transport}

In transport of pencil beams in the air from height $z_0$ to the next device at height $z_1$, the positions and the variances, $\overline{x}_{ij}$, $\overline{y}_{ij}$, $\overline{\theta t}_{ij}$, and $\overline{t^2}_{ij}$, are modified by
\begin{eqnarray}
\fl
\Delta \overline{x}_{ij} = \overline{a_\mathrm{x}}_{ij}\, \Delta z, \quad
\Delta \overline{y}_{ij} = \overline{a_\mathrm{y}}_{ij}\, \Delta z, \quad
\Delta z = z_1-z_0, \qquad
\frac{\Delta s}{\Delta z} = -\sqrt{\overline{a_\mathrm{x}}_{ij}^2+\overline{a_\mathrm{y}}_{ij}^2+1}, \nonumber\\*
\fl
\Delta \overline{\theta t}_{ij} = \overline{\theta^2}_{ij}\, \Delta s, \qquad
\Delta \overline{t^2}_{ij} = \left(2\, \overline{\theta t}_{ij} + \overline{\theta^2}_{ij}\, \Delta s \right) \Delta s, 
\end{eqnarray}
according to the formulation in \sref{sec_physics}. 
We denote here the height immediately before the device as $^{\sharp}\!z_1$, where the effects of the device have not yet been included, and the PB parameters are modified in the transport as $\overline{t^2}(^{\sharp}\!z_1)_{ij} = \overline{t^2}(z_0)_{ij} + \Delta \overline{t^2}_{ij}$, for example.

\subsubsection{Regeneration }\label{sec_regeneration}

Each of the transported pencil beams at height $^{\sharp}\!z_1$ immediately before the device contains particles whose probability density in slopes and positions $(a_\mathrm{x},a_\mathrm{y}, x, y)$ is described by phase-space distribution
\begin{eqnarray}
\fl F_{ij}(a_\mathrm{x}, a_\mathrm{y}, x, y) = \frac{N_{ij}}{4\, \pi^2}\,\left(\overline{\theta^2}_{ij}\,\overline{t^2}_{ij}-\overline{\theta t}_{ij}^2\right)^{-1}\, \rme^{-\frac{(a_\mathrm{x}-\overline{a_\mathrm{x}}_{ij})^2 +(a_\mathrm{y}-\overline{a_\mathrm{y}}_{ij})^2}{\overline{\theta^2}_{ij} \overline{t^2}_{ij}-\overline{\theta t}_{ij}^2}\,\frac{\overline{t^2}_{ij}}{2}}\nonumber\\*
\times \rme^{-\frac{(a_\mathrm{x}-\overline{a_\mathrm{x}}_{ij}) (x-\overline{x}_{ij})+(a_\mathrm{y}-\overline{a_\mathrm{y}}_{ij}) (y-\overline{y}_{ij})}{\overline{\theta^2}_{ij} \overline{t^2}_{ij}-\overline{\theta t}_{ij}^2}\, \overline{\theta t}_{ij}}\,\rme^{-\frac{(x-\overline{x}_{ij})^2+(y-\overline{y}_{ij})^2}{\overline{\theta^2}_{ij} \overline{t^2}_{ij}-\overline{\theta t}_{ij}^2} \frac{\overline{\theta^2}_{ij}}{2}},
\end{eqnarray}
which is derived from two-dimensional extension of \eref{eq_3} with approximation $\theta \approx \tan \theta \approx (\overline{a}_{ij}-a)$.
In the absence of residual range variation, these particles are indistinguishable among the pencil beams and are superposed to form a broad beam with fluence, mean slopes, and angular variance of the particles,
\begin{eqnarray}
\fl
\Phi(x,y) = \int_{-\infty}^{\infty}\!\!\!\! \rmd a_\mathrm{x} \int_{-\infty}^{\infty}\!\!\!\! \rmd a_\mathrm{y} \sum_{i,j} F_{ij}(a_\mathrm{x}, a_\mathrm{y}, x, y) = \sum_{i,j} \frac{N_{ij}}{2\, \pi\, \overline{t^2}_{ij}}\, \rme^{-
\frac{\left(x-\overline{x}_{ij}\right)^2 + \left(y-\overline{y}_{ij}\right)^2} {2\, \overline{t^2}_{ij}}}, \label{eq_12}
\\ 
\fl
\overline{a_\mathrm{x}}(x,y) = \frac{1}{\Phi(x,y)} \int_{-\infty}^{\infty}\!\!\!\! \rmd a_\mathrm{x} \int_{-\infty}^{\infty}\!\!\!\! \rmd a_\mathrm{y} \sum_{i,j} a_\mathrm{x}\, F_{ij}(a_\mathrm{x}, a_\mathrm{y}, x, y) \nonumber\\*
= \frac{1}{\Phi(x,y)} \sum_{i,j} \frac{N_{ij}}{2\, \pi\, \overline{t^2}_{ij}}\, \rme^{-
\frac{\left(x-\overline{x}_{ij}\right)^2 + \left(y-\overline{y}_{ij}\right)^2} {2\, \overline{t^2}_{ij}}} \left( \overline{a_\mathrm{x}}_{ij}-\frac{x-\overline{x}_{ij}}{\overline{t^2}_{ij}}\, \overline{\theta t}_{ij}\right),
\\
\fl
\overline{a_\mathrm{y}}(x,y) = \frac{1}{\Phi(x,y)} \int_{-\infty}^{\infty}\!\!\!\! \rmd a_\mathrm{x} \int_{-\infty}^{\infty}\!\!\!\! \rmd a_\mathrm{y} \sum_{i,j} a_\mathrm{y}\, F_{ij}(a_\mathrm{x}, a_\mathrm{y}, x, y) \nonumber\\*
= \frac{1}{\Phi(x,y)} \sum_{i,j} \frac{N_{ij}}{2\, \pi\, \overline{t^2}_{ij}}\, \rme^{-
\frac{\left(x-\overline{x}_{ij}\right)^2 + \left(y-\overline{y}_{ij}\right)^2} {2\, \overline{t^2}_{ij}}} \left( \overline{a_\mathrm{y}}_{ij}-\frac{y-\overline{y}_{ij}}{\overline{t^2}_{ij}}\, \overline{\theta t}_{ij} \right),
\\
\fl
\overline{\theta^2}(x,y) \approx \frac{1}{\Phi(x,y)} \int_{-\infty}^{\infty}\!\!\!\! \rmd a_\mathrm{x} \int_{-\infty}^{\infty}\!\!\!\! \rmd a_\mathrm{y} \sum_{i,j} \frac{a_\mathrm{x}^2+a_\mathrm{y}^2}{2}\, F_{ij}(a_\mathrm{x}, a_\mathrm{y}, x, y) -\frac{\overline{a_\mathrm{x}}^2(x,y)+\overline{a_\mathrm{y}}^2(x,y)}{2} \nonumber\\*
= \frac{1}{\Phi(x,y)} \sum_{i,j} \frac{N_{ij}}{2\, \pi\, \overline{t^2}_{ij}}\, \rme^{-
\frac{\left(x-\overline{x}_{ij}\right)^2 + \left(y-\overline{y}_{ij}\right)^2} {2\, \overline{t^2}_{ij}}} \Bigg[
\overline{\theta^2}_{ij}-\frac{\overline{\theta t}_{ij}^2}{\overline{t^2}_{ij}}
\nonumber\\*
+\frac{1}{2} \left( \overline{a_\mathrm{x}}_{ij}-\frac{x-\overline{x}_{ij}}{\overline{t^2}_{ij}} \,\overline{\theta t}_{ij} \right)^2 +\frac{1}{2}\left( \overline{a_\mathrm{y}}_{ij}-\frac{y-\overline{y}_{ij}}{\overline{t^2}_{ij}} \, \overline{\theta t}_{ij} \right)^2 \Bigg]
\nonumber\\*
-\frac{\overline{a_\mathrm{x}}^2(x,y)+\overline{a_\mathrm{y}}^2(x,y)}{2},
\end{eqnarray}
as a function of position $(x,y)$ on the $^{\sharp}\!z_1$ plane, respectively. 
The reconstructed broad beam is redivided into new pencil beams at reinitialized grids on the $z_1$ plane as
\begin{eqnarray}
\fl
\overline{x}(z_1)_{ij} = {x_1}_j, \qquad \overline{y}(z_1)_{ij} = {y_1}_i, \qquad
\frac{\delta_1}{\delta_\mathrm{iso}} = \frac{{x_1}_j}{{x_\mathrm{iso}}_j} = \frac{{y_1}_i}{{y_\mathrm{iso}}_i} = \frac{z_1-z_\mathrm{src}}{z_\mathrm{iso}-z_\mathrm{src}}, \nonumber\\*
\fl
\overline{t^2}(z_1)_{ij} = \frac{\delta_1^2}{12}, \qquad
\overline{\theta t}(z_1)_{ij} = \frac{\delta_1^2}{12\, |z_1-z_\mathrm{src}|}, \qquad
R(z_1)_{ij} = R(z_0)_{ij}.
\end{eqnarray}
Denoting any parameter $p$ at the height immediately before the device as $p^{\sharp} = p(^{\sharp}\!z_1)$, the statistical parameters of the regenerated pencil beams are redefined as
\begin{eqnarray}
\fl
N(z_1)_{ij} = \int_{{x_1}_j-\frac{\delta_1}{2}}^{{x_1}_j+\frac{\delta_1}{2}}\!\!\!\!\rmd x
\int_{{y_1}_i-\frac{\delta_1}{2}}^{{y_1}_i+\frac{\delta_1}{2}}\!\!\!\!\rmd y \, \Phi(x, y;\, ^{\sharp}\!z_1)
\nonumber\\*
= \sum_{k,l} N_{kl}^{\sharp}\, \frac{1}{2} \left[ \mathrm{erf}\!\left(
\case{{x_1}_j-\overline{x}_{kl}^{\sharp}+\frac{\delta_1}{2}}
{\sqrt{2\,\overline{t^2}_{kl}^{\sharp}}}\right) -\mathrm{erf}\!\left(
\case{{x_1}_j-\overline{x}_{kl}^{\sharp} -\frac{\delta_1}{2}}
{\sqrt{2\,\overline{t^2}_{kl}^{\sharp}}}\right) \right]
\nonumber\\*
\times \frac{1}{2} \left[ \mathrm{erf}\!\left(
\case{{y_1}_i-\overline{y}_{kl}^{\sharp} +\frac{\delta_1}{2}}
{\sqrt{2\,\overline{t^2}_{kl}^{\sharp}}}\right) -\mathrm{erf}\!\left(
\case{{y_1}_i-\overline{y}_{kl}^{\sharp} -\frac{\delta_1}{2}}
{\sqrt{2\,\overline{t^2}_{kl}^{\sharp}}}\right) \right],
\\
\fl 
\overline{a_\mathrm{x}}(z_1)_{ij} = \frac{1}{N(z_1)_{ij}}
\int_{{x_1}_j-\frac{\delta_1}{2}}^{{x_1}_j+\frac{\delta_1}{2}}\!\!\!\!\rmd x
\int_{{y_1}_i-\frac{\delta_1}{2}}^{{y_1}_i+\frac{\delta_1}{2}}\!\!\!\!\rmd y \, \overline{a_\mathrm{x}}(x,y;\, ^{\sharp}\!z_1) \, \Phi(x, y;\, ^{\sharp}\!z_1)
\nonumber\\*
= \frac{1}{N(z_1)_{ij}} \sum_{k,l} N_{kl}^{\sharp}\, \frac{1}{2} \left[ \mathrm{erf}\!\left(
\case{{y_1}_i-\overline{y}_{kl}^{\sharp} +\frac{\delta_1}{2}}
{\sqrt{2\,\overline{t^2}_{kl}^{\sharp}}}\right) -\mathrm{erf}\!\left(
\case{{y_1}_i-\overline{y}_{kl}^{\sharp} -\frac{\delta_1}{2}}
{\sqrt{2\,\overline{t^2}_{kl}^{\sharp}}}\right) \right]
\nonumber\\*
\times \Bigg\{ \frac{1}{2} \left[ \mathrm{erf}\!\left(
\case{{x_1}_j-\overline{x}_{kl}^{\sharp}+\frac{\delta_1}{2}}
{\sqrt{2\,\overline{t^2}_{kl}^{\sharp}}}\right) -\mathrm{erf}\!\left(
\case{{x_1}_j-\overline{x}_{kl}^{\sharp} -\frac{\delta_1}{2}}
{\sqrt{2\,\overline{t^2}_{kl}^{\sharp}}}\right) \right] \overline{a_\mathrm{x}}_{kl}^{\sharp} 
\nonumber\\*
-\sqrt{\frac{2}{\pi\, \overline{t^2}_{kl}^{\sharp}}}\, \overline{\theta t}_{kl}^{\sharp}\, \rme^{-\frac{\left({x_1}_j-\overline{x}_{kl}^{\sharp} \right)^2+\frac{\delta_1^2}{4}}{2\, \overline{t^2}_{kl}^{\sharp}}} \sinh\!\left(\case{{x_1}_j-\overline{x}_{kl}^{\sharp}}{2\, \overline{t^2}_{kl}^{\sharp}}\,\delta_1 \right) \Bigg\},
\\
\fl 
\overline{a_\mathrm{y}}(z_1)_{ij} = \frac{1}{N(z_1)_{ij}}
\int_{{x_1}_j-\frac{\delta_1}{2}}^{{x_1}_j+\frac{\delta_1}{2}}\!\!\!\!\rmd x
\int_{{y_1}_i-\frac{\delta_1}{2}}^{{y_1}_i+\frac{\delta_1}{2}}\!\!\!\!\rmd y \, \overline{a_\mathrm{y}}(x,y;\, ^{\sharp}\!z_1) \, \Phi(x, y;\, ^{\sharp}\!z_1)
\nonumber\\*
= \frac{1}{N(z_1)_{ij}} \sum_{k,l} N_{kl}^{\sharp}\, \frac{1}{2} \left[ \mathrm{erf}\!\left(
\case{{x_1}_j-\overline{x}_{kl}^{\sharp} +\frac{\delta_1}{2}}
{\sqrt{2\,\overline{t^2}_{kl}^{\sharp}}}\right) -\mathrm{erf}\!\left(
\case{{x_1}_j-\overline{x}_{kl}^{\sharp} -\frac{\delta_1}{2}}
{\sqrt{2\,\overline{t^2}_{kl}^{\sharp}}}\right) \right]
\nonumber\\* 
\times \Bigg\{ \frac{1}{2} \left[ \mathrm{erf}\!\left(
\case{{y_1}_i-\overline{y}_{kl}^{\sharp} +\frac{\delta_1}{2}}
{\sqrt{2\,\overline{t^2}_{kl}^{\sharp}}}\right) -\mathrm{erf}\!\left(
\case{{y_1}_i-\overline{y}_{kl}^{\sharp} -\frac{\delta_1}{2}}
{\sqrt{2\,\overline{t^2}_{kl}^{\sharp}}}\right) \right] \overline{a_\mathrm{y}}_{kl}^{\sharp}
\nonumber\\*
-\sqrt{\frac{2}{\pi\, \overline{t^2}_{kl}^{\sharp}}}\, \overline{\theta t}_{kl}^{\sharp}\, \rme^{-\frac{\left({y_1}_i-\overline{y}_{kl}^{\sharp} \right)^2+\frac{\delta_1^2}{4}}{2\, \overline{t^2}_{kl}^{\sharp}}} \sinh\!\left(\case{{y_1}_i-\overline{y}_{kl}^{\sharp}}{2\, \overline{t^2}_{kl}^{\sharp}}\,\delta_1 \right) \Bigg\},
\\
\fl 
\overline{\theta^2}(z_1)_{ij} = \frac{1}{N(z_1)_{ij}}
\int_{{x_1}_j-\frac{\delta_1}{2}}^{{x_1}_j+\frac{\delta_1}{2}}\!\!\!\!\rmd x
\int_{{y_1}_i-\frac{\delta_1}{2}}^{{y_1}_i+\frac{\delta_1}{2}}\!\!\!\!\rmd y \, \overline{\theta^2}(x,y;\, ^{\sharp}\!z_1) \, \Phi(x, y;\, ^{\sharp}\!z_1)
\nonumber\\*
\approx \frac{1}{N(z_1)_{ij}} \sum_{k,l} N_{kl}^{\sharp} \Bigg\{
\frac{1}{2} \left[ \mathrm{erf}\!\left(
\case{{x_1}_j-\overline{x}_{kl}^{\sharp}+\frac{\delta_1}{2}}
{\sqrt{2\,\overline{t^2}_{kl}^{\sharp}}}\right) -\mathrm{erf}\!\left(
\case{{x_1}_j-\overline{x}_{kl}^{\sharp} -\frac{\delta_1}{2}}
{\sqrt{2\,\overline{t^2}_{kl}^{\sharp}}}\right) \right] 
\nonumber\\* 
\times \frac{1}{2} \left[ \mathrm{erf}\!\left(
\case{{y_1}_i-\overline{y}_{kl}^{\sharp} +\frac{\delta_1}{2}}
{\sqrt{2\,\overline{t^2}_{kl}^{\sharp}}}\right) -\mathrm{erf}\!\left(
\case{{y_1}_i-\overline{y}_{kl}^{\sharp} -\frac{\delta_1}{2}}
{\sqrt{2\,\overline{t^2}_{kl}^{\sharp}}}\right) \right] \left(\overline{\theta^2}_{kl}^{\sharp} +\frac{{\overline{a_\mathrm{x}}_{kl}^{\sharp}}^2+{\overline{a_\mathrm{y}}_{kl}^{\sharp}}^2}{2} \right) 
\nonumber\\*
- \frac{\overline{\theta t}_{kl}^{\sharp}}{\sqrt{2\, \pi \, \overline{t^2}_{kl}^{\sharp}}} 
\, \rme^{-\frac{\left({x_1}_j-\overline{x}_{kl}^{\sharp} \right)^2+\frac{\delta_1^2}{4}}{2\, \overline{t^2}_{kl}^{\sharp}}}
\Bigg[ \left( \overline{a_\mathrm{x}}_{kl}^{\sharp} -\case{{x_1}_j-\overline{x}_{kl}^{\sharp}}{2\, \overline{t^2}_{kl}^{\sharp}} \overline{\theta t}_{kl}^{\sharp} \right) \sinh\!\left(\case{{x_1}_j-\overline{x}_{kl}^{\sharp}}{2\, \overline{t^2}_{kl}^{\sharp}}\,\delta_1 \right)
\nonumber\\* 
+\frac{\overline{\theta t}_{kl}^{\sharp}}{\overline{t^2}_{kl}^{\sharp}}\,\frac{\delta_1}{4}\, \cosh\!\left(\case{{x_1}_j-\overline{x}_{kl}^{\sharp}}{2\, \overline{t^2}_{kl}^{\sharp}}\,\delta_1 \right) \Bigg] 
\left[ \mathrm{erf}\!\left(
\case{{y_1}_i-\overline{y}_{kl}^{\sharp} +\frac{\delta_1}{2}}
{\sqrt{2\,\overline{t^2}_{kl}^{\sharp} }}\right) -\mathrm{erf}\!\left(
\case{{y_1}_i-\overline{y}_{kl}^{\sharp} -\frac{\delta_1}{2}}
{\sqrt{2\,\overline{t^2}_{kl}^{\sharp}}}\right) \right]
\nonumber\\*
- \frac{\overline{\theta t}_{kl}^{\sharp}}{\sqrt{2\, \pi \, \overline{t^2}_{kl}^{\sharp}}}
\, \rme^{-\frac{\left({y_1}_i-\overline{y}_{kl}^{\sharp} \right)^2+\frac{\delta_1^2}{4}}{2\, \overline{t^2}_{kl}^{\sharp}}}
\Bigg[ \left(\overline{a_\mathrm{y}}_{kl}^{\sharp} -\case{{y_1}_i-\overline{y}_{kl}^{\sharp}}{2\, \overline{t^2}_{kl}^{\sharp}} \overline{\theta t}_{kl}^{\sharp} \right) \sinh\!\left(\case{{y_1}_i-\overline{y}_{kl}^{\sharp}}{2\, \overline{t^2}_{kl}^{\sharp}}\,\delta_1 \right)
\nonumber\\*
+\frac{\overline{\theta t}_{kl}^{\sharp}}{\overline{t^2}_{kl}^{\sharp}}\,\frac{\delta_1}{4}\, \cosh\!\left(\case{{y_1}_i-\overline{y}_{kl}^{\sharp}}{2\, \overline{t^2}_{kl}^{\sharp}}\,\delta_1 \right) \Bigg] \left[ \mathrm{erf}\!\left(
\case{{x_1}_j-\overline{x}_{kl}^{\sharp} +\frac{\delta_1}{2}}
{\sqrt{2\,\overline{t^2}_{kl}^{\sharp} }}\right) -\mathrm{erf}\!\left(
\case{{x_1}_j-\overline{x}_{kl}^{\sharp} -\frac{\delta_1}{2}}
{\sqrt{2\,\overline{t^2}_{kl}^{\sharp}}}\right) \right]
\nonumber\\*
 \Bigg\} -\frac{\overline{a_\mathrm{x}}(z_1)_{ij}^2 + \overline{a_\mathrm{y}}(z_1)_{ij}^2}{2},
\end{eqnarray}
where the exponential, error, and hyperbolic functions will be quickly enumerated with the standard math library.

\subsubsection{Collimation }\label{sec_collimation}

A collimator is described as a matrix of transmission factors, $\bi{T}$, with the same grids as those for the regenerated pencil beams at its downstream face ignoring the collimator thickness (Kanematsu \etal 2006).
For the pencil beams transported to and regenerated at $z_1$, the collimation modifies the relative number of particles immediately after the collimator at height $^{\flat}\!z_1$, 
\begin{eqnarray}
N(^{\flat}\!z_1)_{ij} = N(z_1)_{ij}\, T_{ij},
\end{eqnarray}
where transmission $T_{ij}$ has value 1 when grid $ij$ is in the collimator aperture or otherwise 0.
A series of the processes of transport, regeneration, and collimation can be repeated for multiple collimators as long as the residual ranges have not been varied in the field.

\subsubsection{Range compensation }

A RCF is a sculptured object designed to absorb extra ranges of the incident particles beyond the target, which also inevitably adds nonuniform scattering that deteriorates particle equilibrium and consequently field uniformity. 
The RCF made of a material of effective density $\rho$ and specific radiation length $X_0$ is assumed to have a flat downstream face located at height $z_1$ and a shape described by a matrix of thicknesses, $\bi{S}$.
Pencil beam $ij$ has a path length of approximately $S_{ij}$ in the RCF, ignoring the small beam-divergence effect.

We first transport the pencil beams to the downstream face of the RCF at $z_1$ ignoring the interactions with matter as described in \sref{sec_transport}, which leads to $\overline{\theta t}(^{\sharp}\!z_1)_{ij}$ and $\overline{t^2}(^{\sharp}\!z_1)_{ij}$.
For the first RCF before which all the pencil beams have the same residual range, we regenerate them as described in \sref{sec_regeneration}. 
Thus we obtain $\overline{\theta^2}(z_1)_{ij}$, $\overline{\theta t}(z_1)_{ij}$, and $\overline{t^2}(z_1)_{ij}$.
On exit from the RCF, the range $R_{ij}$ and the angular variance $\overline{\theta^2}_{ij}$ are modified by
\begin{eqnarray}
\fl \Delta R_{ij} = -\rho\, S_{ij}, \qquad 
\Delta \overline{\theta^2}_{ij} = \left[1+\frac{1}{9}\, \log_{10}\!\left( \frac{S_{ij}}{X_0}\right) \right]^2 \left( \frac{14.1\, Z}{\overline{pv}_{ij}}\right)^2 \frac{S_{ij}}{X_0},
\end{eqnarray}
where $Z$ and $\overline{pv}_{ij}$ are the charge and the mean momentum-velocity product of the particles in beam $ij$.
The growths in spatial variance and covariance for the travel from the effective scattering point that is approximated to the midpoint,
\begin{eqnarray}
\fl \Delta \overline{\theta t}_{ij} = \Delta \overline{\theta^2}_{ij} \left(0.5\,S_{ij}\right), \qquad
\Delta \overline{t^2}_{ij} = \Delta \overline{\theta^2}_{ij} \left(0.5\,S_{ij}\right)^2,
\end{eqnarray}
are correctively added to $\overline{\theta t}(z_1)_{ij}$ and $\overline{t^2}(z_1)_{ij}$ for the pencil beams exiting from the RCF at $^{\flat}\!z_1$.
Note that this scattering correction along with the in-air transport is mathematically equivalent to a sequence of transport to the effective scattering point, effective-point scattering, and transport to the RCF reference face.

\subsubsection{Edge sharpening }

When a PTC is placed downstream of a RCF, a sharp-edged field must be formed by the PTC, whereas the modulated range loss and scattering effects originated by the RCF prevents from applying the regeneration technique.
For an approximate solution, we partly modify the transported and collimated pencil beams near the collimator edge at height $z_1$ in such a way that the spreads of the outgoing pencil beams at height $^{\flat}\!z_1$ are conditionally scaled to the distances of closest approach to the collimator edge, $d_{ij}$, with control parameter $\alpha$,
\begin{eqnarray}
\fl \overline{t^2}(^{\flat}\!z_1)_{ij} = \min \left( \frac{d_{ij}^2}{\alpha^2},\ \overline{t^2}(z_1)_{ij} \right), \qquad
\overline{\theta t}(^{\flat}\!z_1)_{ij} = \frac{ \overline{t^2}(^{\flat}\!z_1)_{ij}}{|z_1-z_\mathrm{src}|},
\\ 
\fl d_{ij}^2 = \min_{\forall (i'j') \in \left(T_{i'j'}=0\right)} \left(\left(\left| {x_1}_{j'}-{x_1}_j\right|-\frac{\delta_1}{2}\right)^2\Big|_{j' \neq j}+\left(\left| {y_1}_{i'}-{y_1}_i\right|-\frac{\delta_1}{2}\right)^2\Big|_{i' \neq i}\right) ,
\end{eqnarray}
where the conditional terms are applicable when the blocked pixel $i'j'$ is in different row or column.
The beam spots are regulated so that the radii of $\alpha$ standard deviations will be within the aperture.

\begin{figure}
\begin{indented}
\item[] \includegraphics[width=10 cm]{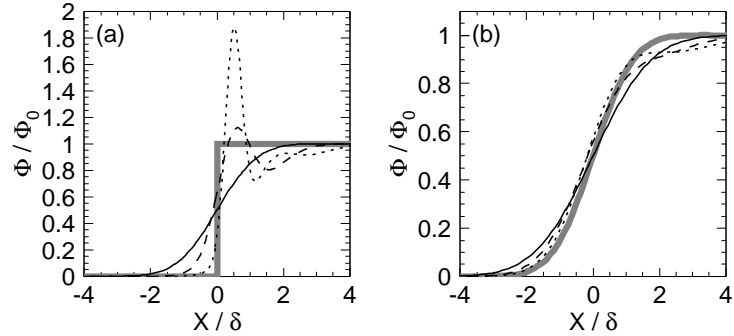}
\end{indented}
\caption{Calculated profiles in fluence $\Phi$ relative to the open beam fluence $\Phi_0$ on the (a) collimator and (b) isocenter planes with lateral position $x$ in units of local PB interval $\delta$.
The thin solid and thick gray lines are the original and ideal ones while the dashed and dotted lines are edge-sharpened ones with $\alpha$ = 1 and 2, respectively.}
\label{fig_2}
\end{figure}

\Fref{fig_2} shows an example with $N(z_1)_{ij} = 1$, $T_{ij} = H({x_1}_{ij})$, $\overline{t^2}(z_1)_{ij} = \delta_1^2$, and $\overline{t^2}(z_\mathrm{iso})_{ij} = \overline{t^2}(z_1)_{ij} +\delta_\mathrm{iso}^2$, where the step function, $H(x) = 1$ for $x > 0$ or otherwise 0, represents a half-field collimation.
In this case, the edge sharpening successfully reduced the spread near the collimator edge though with strong distortion.
In penumbra behavior on the isocenter plane, the edge-sharpened profiles agreed well with the ideal one, while the unwanted distortions mostly collapsed in the transport.
The distortions should naturally have spatial structure as small as the spread being sharpened, which is normally much smaller than the subsequent spread in the patient, and will generally collapse in dose distributions.

\subsection{Implementation and validation}

\subsubsection{Experimental apparatus }

\begin{figure}
\begin{indented}
\item[] \includegraphics[width=10 cm]{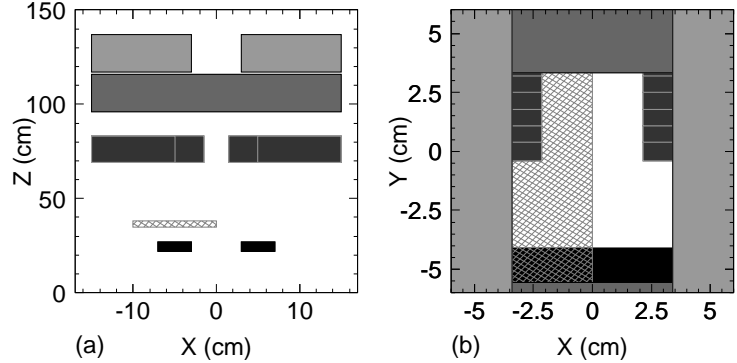}
\end{indented}
\caption{Illustration of the modeled beam-customization devices in (a) side view and (b) beam's eye view, where the filled areas represent XJC, YJC, MLC, and PTC from upstream to downstream, and the hatched area represents RCF.}
\label{fig_3}
\end{figure}

We examine here how well the formulated beam-customization model can handle combinational collimation with one of the therapeutic beam lines of accelerator facility HIMAC at National Institute of Radiological Sciences (Kanai \etal 1999).
A $^{12}$C$^{6+}$ beam of per-nucleon kinetic energy $E/A = 350$ MeV was broadened with the wobbler-scatterer system to form a 15-cm$\phi$ uniform (nominally $\pm 2.5\%$) field with a source at height 950 cm from the isocenter. 
The resultant residual range 19.6 cm in water corresponds to $E/A = 327$ MeV.
 As shown in \fref{fig_3}, the beam was customized with an XJC at 117 cm, a YJC at 96 cm, a MLC with 23 pairs of leaves at 69 cm, a RCF at 35 cm, and a PTC at 22 cm in height.
The XJC and YJC apertures were set to $(-3.0,+3.0)$ cm in $x$ and $(-5.0,+3.0)$ cm in $y$, respectively, and the 11 lower and 12 upper leaf pairs were set to $(-5.0,+5.0)$ cm and $(-2.0,+2.0)$ cm in $x$, respectively.
A 3-cm thick PMMA plate that only covered the left half of the field and an 8-cm-square aperture block were placed as the RCF and the PTC, respectively.

\subsubsection{Measurement }

Dose profiles in the air on the isocenter plane were measured with a 15-mm$^3$ 2-mm$\phi$ pinpoint ionization chamber at intervals of 1 to 2 mm with movement precision of 0.1 mm.
The profiling axes were in $x$ at $y = -2.0$ cm and $+1.0$ cm and in $y$ at $x = -1.0$ cm and $+1.0$ cm with 0.5-mm alignment precision, where the field edges were approximately formed by the individual collimators with and without the RCF.
The measured doses were then divided by the corresponding open beam doses $D_0$ without collimation nor compensation to correct the non-uniformity of the broad beam.
The penumbra sizes are derived from the dose-ratio profiles with dosimeter-size correction, $d_\mathrm{20\to80} = \surd({d'}_\mathrm{20\to80}^2-1.68^2\,\sigma_\mathrm{dos}^2)$, where ${d'}_\mathrm{20\to80}$ is the observed 20\%--80\% distance and $\sigma_\mathrm{dos} = 0.5$ mm is the geometrically estimated rms dosimeter size for the 2-mm$\phi$ diameter. 
For dosimetric analysis, the tissue-air ratio for the 3-cm PMMA plate was measured with the same beam to be 0.951.

\subsubsection{Analytic estimation }

In the local broad-beam approximation near the individual collimator edges, penumbra behaviors in relative fluence with respect to the open beam fluence, $\Phi/\Phi_0$, are calculated as
\begin{eqnarray}
\frac{\Phi}{\Phi_0} = \frac{1}{2}+\frac{1}{2}\, \mathrm{erf}\left(\frac{d_\mathrm{min}}{\sqrt 2\, \sigma_\mathrm{tot}}\right),
\end{eqnarray}
where the signed distance to the closest point on the collimator edge,  $d_\mathrm{min}$,  is positive (negative) in (out of) the field (Hong \etal 1996) and the total rms spread $\sigma_\mathrm{tot}$ is related to the penumbra size by $d_\mathrm{20\to80} = 1.68\,\sigma_\mathrm{tot}$.

The source size is estimated to be $\sigma_\mathrm{src} = 2.54$~cm inversely from \eref{eq_1} with the measured penumbra size $d_\mathrm{20\to80} = 4.8$~mm at the upper $y$ edge at $x = +1$~cm formed by the YJC with $|z-z_\mathrm{col}| = 96$ cm and $\sigma_\mathrm{rcf} = \sigma_\mathrm{pt} = 0$.
The 3-cm PMMA plate at the RCF adds angular spread $\theta_\mathrm{rcf} = \sigma_\mathrm{rcf}/|z-z_\mathrm{rcf}| = \surd(\Delta \overline{\theta^2}) = 3.3$ mrad from \eref{eq_6} and \eref{eq_7} with effective density $\rho = 1.16$ and specific radiation length $X_0 = 34.07$ cm (Yao \etal 2006).

\Eref{eq_1} leads to the penumbra sizes for the two left edges in the $x$ profiles and the upper edge in the $y$ profile at $x = -1$ cm, where the RCF is downstream of all the active collimators, \eref{eq_2} does for the lower $y$ edge at $x = -1$ cm, where the PTC is downstream of the RCF, and both \eref{eq_1} and \eref{eq_2} equivalently do for the edges in the right half field with no actual RCF.
In dosimetric analysis, the tissue-air ratio 0.951 for the PMMA is multiplied to the analytic fluences in the $x < 0$ region.

\subsubsection{Model computation }

In the framework described in \sref{sec_development}, pencil beams were generated at the XJC, immediately collimated, transported to the YJC, regenerated, collimated, transported to the MLC, regenerated, collimated, transported to the RCF, regenerated, range-compensated, transported to PTC, collimated, edge-sharpened, and transported to the isocenter plane, where the grid spacing on the isocenter plane and the edge-sharpening parameter were chosen to be $\delta_\mathrm{iso} = 0.1$ cm and $\alpha = 2$ in the model computation in addition to the common parameters in the analytic estimation. 

The fluence distributions at the devices were calculated with \eref{eq_12} and similarly the in-air dose distribution on the isocenter plane was calculated with
\begin{eqnarray}
D(x,y) = \sum_{i,j} \frac{N_{ij}\, {D_\mathrm{BB}}_{ij}}{2\, \pi\, \overline{t^2}_{ij}}\, \rme^{-
\frac{\left(x-\overline{x}_{ij}\right)^2 + \left(y-\overline{y}_{ij}\right)^2} {2\, \overline{t^2}_{ij}}},
\end{eqnarray}
where ${D_\mathrm{BB}}_{ij}$ is the tissue-air ratio amounting to 0.951 for pencil beam $ij$ with $\overline{x}_{ij} < 0$ or otherwise 1. 

\section{Results}

\subsection{Field-formation process}

\begin{figure}
\indented{\item[]\includegraphics[width=6 cm]{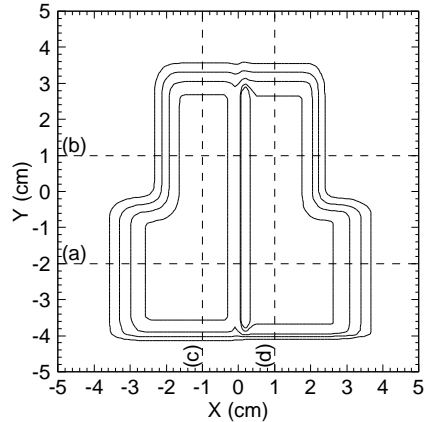}}
\caption{Computed fluence distribution on the isocenter plane, where the iso-fluence lines are 20\%, 50\%, 80\%, 97.5\%, and 102.5\% and the dashed lines indicate the profiling locations with figure-part symbols in \fref{fig_5}.}
\label{fig_4}
\end{figure}

\Fref{fig_4} shows the fluence distribution on the isocenter plane in the model computation, which are consistent with expectations such that the field edges formed at the collimators become gentler with the traveling distance.
The dip and the bump around $x \approx 0$ are due to lateral particle disequilibrium caused by the the PMMA half plate at the RCF.
The whole computation took only about ten seconds with FORTRAN interpreter in analysis package PAW by CERN on PowerPC G5 2-GHz processor by Apple/IBM. 
The short computational time is a great advantage of the deterministic calculation compared to Monte Carlo simulations which may need orders of magnitude more time (Paganetti \etal 2004).

\subsection{Dose-profile analysis}

\begin{figure}
\includegraphics[width=13cm]{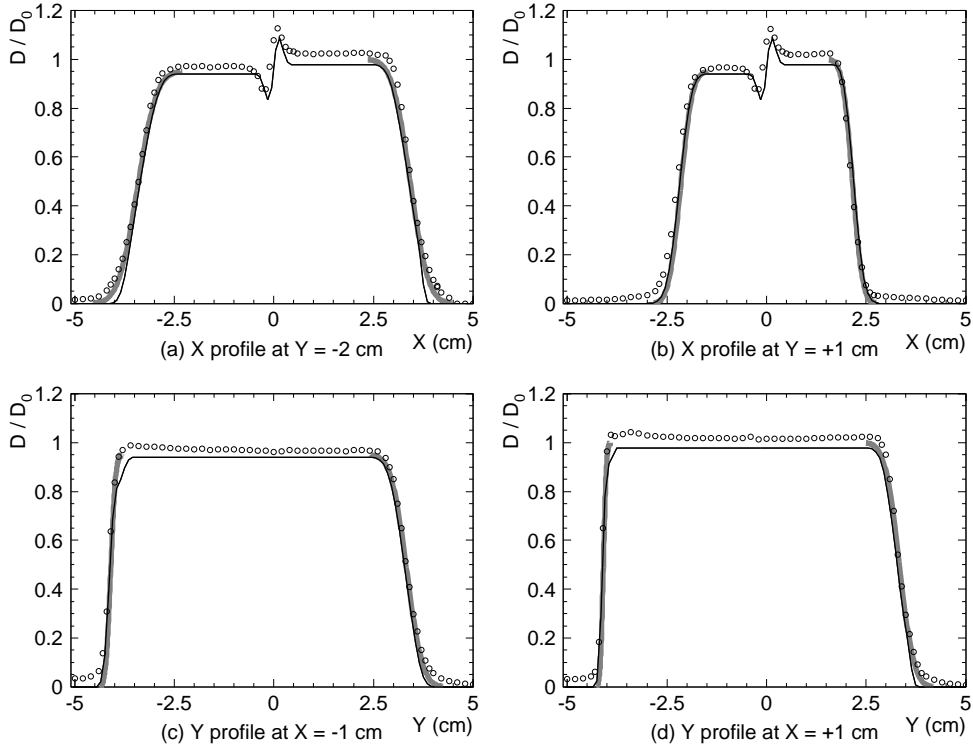}
\caption{Profiles in dose $D$ relative to the open beam dose $D_0$, along $x$ axis at (a) $y = -2$ cm, (b) $y = +1$ cm, along $y$ axis at (c) $x = -2$ cm, and (d) $x = +1$ cm, where the black lines, the thick gray lines, and the open circles are the model-computed, analytic, and measured ones, respectively.}
\label{fig_5}
\end{figure}

The profiles of dose ratios between the customized and open beams, $D/D_0$, are plotted in \fref{fig_5}, where non-uniformity of the actual broad beam should have been compensated.
The measured profiles involve spatial uncertainties from the grid interpolation (0.2~mm) and the global misalignment (0.5~mm) and dose uncertainty of 0.3\% from the dosimeter resolution.

\subsubsection{Penumbra }

\begin{table}
\caption{Experimentally measured, analytically estimated, and model-computed penumbra sizes in the dose profiles shown in \fref{fig_5}.}
\begin{indented}
\item[]
\begin{tabular}{l l l l l l}
\br
Profiling & Interested & Effective & \multicolumn{3}{l}{20\%--80\% penumbra size (mm)} \\
\cline{4-6}
position & edge side & device(s) & measured & analytic & computed \\
\mr
$y=-2$ cm & $x$ left & XJC+RCF & 6.4 & 6.3 & 6.0 \\
$y=-2$ cm & $x$ right & XJC & 5.8 & 6.0 & 5.9 \\
$y=+1$ cm & $x$ left & MLC+RCF & 4.6 & 3.9 & 4.1 \\
$y=+1$ cm & $x$ right & MLC & 3.7 & 3.3 & 3.5 \\
$x=-1$ cm & $y$ lower & PTC+RCF & 2.3 & 1.6 & 2.3 \\
$x=-1$ cm & $y$ upper & YJC+RCF & 5.6 & 5.2 & 5.2 \\
$x=+1$ cm & $y$ lower & PTC & 1.4 & 1.0 & 1.5 \\
$x=+1$ cm & $y$ upper & YJC & 4.8$^{\rm a}$ & 4.8$^{\rm a}$ & 5.0 \\
\br
\end{tabular}
\item[] $^{\rm a}$ The YJC penumbras were calibrated to determine the source size in the model.
\end{indented}
\label{tab_1}
\end{table}

\Tref{tab_1} summarizes the resultant 20\%--80\% penumbra sizes, where the measured ones involve the dosimeter-size correction with the estimated uncertainty of 0.2 mm.
The discrepancies among the measured, analytic, and computed ones turned out to be at a submillimeter level and are consistent with the uncertainties in the experimental and theoretical systems, excluding the global misalignment that should not be influential to the penumbra sizes.

\subsubsection{Collimator scatter }

In dose ratios between the customized beam and the open beam, the measured ones turned out to be larger than the computed counterparts by about 2\% throughout the field.
This irreproducible dose excess may have come from particles hard-scattered by the collimators (Kusano \etal 2007a) ignored in the model.
The collimator-scatter contribution should naturally attenuate with depth, which is consistent with the observation such that the excess was smaller with the PMMA half plate at RCF in the $x < 0$ region in \fref{fig_5}.

\subsubsection{Scatter modulation }

Scatter by the PMMA half plate caused a dip and a bump at $x \approx 0$ in \fref{fig_5}(a) and \fref{fig_5}(b), where the computed bump/dip ratios were (a) 1.30 and (b) 1.30 while the measured counterparts were (a) 1.28 and (b) 1.28, respectively.
Considering the dosimetric limitations, that may be an excellent agreement and indicates accurate evaluation of the RCF scatter in the field, which the conventional PB algorithms must fail to achieve.

\subsubsection{Multiplicative collimation }

In the model computation, the PTC aperture, $|x|, |y| < 4.09$ cm on the isocenter plane, naturally influenced the field edges formed by the XJC at $x = \pm 3.42$ cm and by the YJC at $y = +3.34$ cm, and suppressed the dose tails in \fref{fig_5}.
Such effect may physically exist, but unfortunately resulted in larger discrepancy from the measured data for the ignored collimator scatter.

\section{Discussion}

Hong \etal (1996) suggested using the height of the nearest collimator for $z_\mathrm{col}$ in \eref{eq_1} for combinational collimation in the PB algorithm in analogy with the broad-beam algorithm.
However, the nearest collimator would suddenly switch from one to the other in the middle of the field, where the PB algorithm would cause artifactual dose fluctuation regardless of heterogeneity.
In contrast, the edge-sharpening process in this work only applies to the field-edge region and therefore there will be no unphysical collimator dependence in the middle of the field.

Though we assumed here a single RCF with a flat downstream face, the formulation is already capable to handle multiple RCFs of any structure as well.
In the absence of a RCF, the regeneration technique should be applied on the entrance to the patient to remove the unphysical collimator-height dependence.
The collimator thickness effect can be handled in this framework with virtual multiplicative collimation with spacing corresponding to the thickness in a manner similar to and more sofisticated than the method by Kanematsu \etal (2006).
The present computational model for beam customization will provide a set of pencil beams to various PB algorithms to calculate dose distributions in treatment planning.

The hard-scattered secondary particles are generally out of the scope of PB algorithms including this work.
Practical modeling of collimator and phantom scatters in nuclear interactions may have yet to be studied (van Luijk \etal 2001, Pedroni \etal 2003, and Kusano \etal 2007b), or could possibly be only resolved by Monte Carlo methods (Paganetti 2002, Kase \etal 2006, and Titt \etal 2008).
Fortunately, the clinically relevant dose in ion-beam therapy is generally dominated by the primary particles (Matsufuji \etal 2003).

The regeneration technique here only considers particle flow in the transverse plane in the absence of heterogeneity.
In contrast, the pencil-beam redefinition algorithm, which is based on the same principle, deals with electron transport in the presence of heterogeneity by introducing additional particle flow in the energy space (Shiu and Hogstrom 1991).
Such extension should be also valid for heavy charged particles and would further improve the accuracy of dose distributions.
However, an order of magnitude smaller scatter and an order of magnitude better spatial accuracy generally required for heavy-charged-particle radiotherapy could possibly make implementation of the energy flow less significant or less practical.

\section{Conclusions}

The phase-space theory for beam transport has been successfully applied to computational modeling of beam-customization devices in heavy-charged-particle radiotherapy to accurately deal with multiple scattering in the presence of multiple collimators and a range-compensating filter in fluence distributions at a submillimeter level.

The present computational model is efficient and readily applicable to pencil-beam algorithms for treatment planning and will enable combinational collimation and compensation to make the best use of the beam-customization devices in clinical practice.

\References

\item[] Akagi T, Kanematsu N, Takatani Y, Sakamoto H, Hishikawa Y and Abe M 2006 Scatter factors in proton therapy with a broad beam \PMB {\bf 51} 1919--28

\item[] Boyd R A, Hogstrom K R and Starkschall G 2001 Electron pencil-beam redefinition algorithm in the presence of heterogeneities {\it Med. Phys.} {\bf 28} 2096--104

\item[] Chi P C, Hogstrom K R, Starkschall G, Antolak J A and Boyd R A 2005 Modeling of skin collimation using the electron pencil beam redefinition algorithm {\it Med. Phys.} {\bf 32} 3409--18

\item[] Deasy J O 1998 A proton dose calculation algorithm for conformal therapy simulations based on Moli\`ere's theory of lateral deflections {\it Med. Phys.} {\bf 25} 476--83

\item[] Eyges L 1948 Multiple scattering with energy loss {\it Phys. Rev.} {\bf 74} 1534--5

\item[] Gottschalk B, Koehler A M, Schneider R J, Sisterson J M and Wagner M S 1993 Multiple Coulomb scattering of 160 MeV protons {\it Nucl. Instrum. Methods B} {\bf 74} 467--90

\item[] Highland V L 1975 Some practical remarks on multiple scattering {\it Nucl. Instrum. Methods} {\bf 129} 497--9

\item[] Hollmark M, Uhrdin, J, Belki\'c D\v{z}, Gudowska I and Brahme A 2004 Influence of multiple scattering and energy loss straggling on the absorbed dose distributions of therapeutic light ion beams: I. Analytical pencil beam model \PMB {\bf 49} 3247--65

\item[] Hong L, Goitein M, Bucciolini M, Comiskey R, Gottschalk B, Rosenthal S, Serago C and Urie M 1996 A pencil beam algorithm for proton dose calculations \PMB {\bf 41} 1305--30

\item[] ICRU-49 1993 Stopping Powers and Ranges for Protons and Alpha Particles {\it International Commission on Radiation Units and Measurements Report} 49 (Bethesda, MD: ICRU)

\item[] IEC-61217 2002-3 Radiotherapy equipment-coordinates, movements, and scales {\it International Standard} IEC 61217 Ed. 1.1 (Geneva: International Electrotechnical Commission)

\item[] J\"{a}kel O, K\"{a}mer M, Karger C P and Debus J 2001 Treatment planning for heavy ion radiotherapy: clinical implementation and application \PMB {\bf 46} 1101--16

\item[] Kanai T \etal 1999 Biophysical characteristics of HIMAC clinical irradiation system for heavy-ion radiation therapy {\it Int. J. Radiat. Oncol. Biol. Phys.} {\bf 44} 201--10

\item[] Kanematsu N, Akagi T, Futami Y, Higashi A, Kanai T, Matsufuji N, Tomura H and Yamashita H 1998 A proton dose calculation code for treatment planning based on the pencil beam algorithm {\it Jpn. J. Med. Phys.} {\bf 18} 88--103

\item[] Kanematsu N, Matsufuji N, Kohno R, Minohara S and Kanai T 2003 A CT calibration method based on the polybinary tissue model for radiotherapy treatment planning \PMB {\bf 48} 1053--64

\item[] Kanematsu N, Akagi T, Takatani Y, Yonai S, Sakamoto H and Yamashita H 2006 Extended collimator model for pencil-beam dose calculation in proton radiotherapy \PMB {\bf 51} 4807--17

\item[] Kanematsu N, Torikoshi M, Mizota M and Kanai T 2007 Secondary range shifting with range compensator for reduction of beam data library in heavy-ion radiotherapy {\it Med. Phys.} {\bf 34} 1907-10

\item[] Kanematsu N, Yonai S and Ishizaki A 2008 The grid-dose-spreading algorithm for dose distribution calculation in heavy charged particle radiotherapy {\it Med. Phys.} {\bf 35} 602--7

\item[] Kase Y, Kanematsu N, Kanai T and Matsufuji N 2006 Biological dose calculation with Monte Carlo physics simulation for heavy-ion radiotherapy \PMB {\bf 51} N467--75

\item[] Kohno R, Kanematsu N, Yusa K, and Kanai T 2004a Experimental evaluation of analytical penumbra calculation model for wobbled beams {\it Med. Phys.} {\bf 31} 1153--7

\item[] Kohno R, Kanematsu N, Kanai T and Yusa K 2004b Evaluation of a pencil beam algorithm for therapeutic carbon ion beam in presence of bolus {\it Med. Phys.} {\bf 31} 2249--53

\item[] Kusano Y, Kanai T, Kase Y, Matsufuji N, Komori M, Kanematsu N, Ito A and Uchida H 2007a Dose contributions from large-angle scattered particles in therapeutic carbon beams {\it Med. Phys.} {\bf 34} 193--8

\item[] Kusano Y, Kanai T, Yonai S, Komori M, Ikeda N, Tachikawa Y, Ito A and Uchida H 2007b Field-size dependence of doses of therapeutic carbon beams {\it Med. Phys.} {\bf 34} 4016--22

\item[] Lomax \etal 2001 Intensity modulated proton therapy: A clinical example {\it Med. Phys.} {\bf 28} 317--24

\item[] Matsufuji N, Fukumura A, Komori M, Kanai T and Kohno T 2003 Influence of fragment reaction of relativistic heavy charged particles on heavy-ion radiotherapy \PMB {\bf 48} 1605--23

\item[] Paganetti H 2002 Nuclear interactions in proton therapy: dose and relative biological effect distributions originating from primary and secondary particles \PMB {\bf 47} 747--64

\item[] Paganetti H, Jiang H, Lee S Y and Kooy H M 2004 Accurate Monte Carlo simulations for nozzle design, commissioning and quality assurance for a proton radiation therapy facility {\it Med. Phys.} {\bf 31} 2107--18

\item[] Pedroni E, Scheib S, B\"ohringer T, Coray A, Grossmann M, Lin S and Lomax A 2005 Experimental characterization and physical modelling of the dose distribution of scanned pencil beams \PMB {\bf 50} 541--61

\item[] Petti P L 1992 Differential-pencil-beam dose calculations for charged particles {\it Med. Phys.} {\bf 19} 137--49

\item[] Russell K R, Isacsson U, Saxner M, Ahnesj\"o A, Montelius E, Grusell E, Vallhagen~Dahlgren C, Lorin S and Glimelius B 2000 Implementation of pencil kernel and depth penetration algorithms for treatment planning of proton beams \PMB {\bf 45} 9--27

\item[] Shiu A S and Hogstrom K R 1991 Pencil-beam redefinition algorithm for electron dose distributions {\it Med. Phys.} {\bf 18} 7--18

\item[] Storchi P R M and Huizenga H 1985 On a numerical approach of the pencil beam model \PMB {\bf 30} 467--73

\item[] Szymanowski H, Mazal A, Nauraye C, Biensan S, Murillo M C, Caneva S, Gaboriaud G and Rosenwald J C 2001 Experimental determination and verification of the parameters used in a proton pencil beam algorithm {\it Med. Phys.} {\bf 28} 975--87

\item[] Tomura H, Kanai T, Higashi A, Futami Y, Matsufuji N, Endo M, Soga F and Kawachi K 1998 Analysis of the penumbra for uniform irradiation fields delivered by a wobbler method {\it Jpn. J. Med. Phys.} {\bf 18} 42--56

\item[] Titt U, Zheng Y, Vassiliev O N and Newhauser W D 2008 Monte Carlo investigation of collimator scatter of proton-therapy beams produced using the passive scattering method \PMB {\bf 53} 487--504

\item[] van Luijk P, van’t Veld A A, Zelle H D and Schippers J M 2001 Collimator scatter and 2D dosimetry in small proton beams {\PMB} {\bf 46} 653--67

\item[] Yao W M, \etal 2006 Review of particle physics {\it J. of Phys. G} {\bf 33} 1--1232

\endrefs

\end{document}